\documentclass[a4paper,11pt]{article}
\input epsf
\newcounter{fixy}
\begin{document}
\newenvironment{fixy}[1]{\setcounter{figure}{#1}}
{\addtocounter{fixy}{1}}
\renewcommand{\thefixy}{\arabic{fixy}}
\renewcommand{\thefigure}{\thefixy\alph{figure}}
\setcounter{fixy}{1}
\baselineskip 100pt
\renewcommand{\arraystretch}{0.666666666}
\large
\parskip.2in
\newcommand{\be}{\begin{equation}}
\newcommand{\ee}{\end{equation}}
\newcommand{\br}{\bar}
\newcommand{\fr}{\frac}
\newcommand{\lm}{\lambda}
\newcommand{\ra}{\rightarrow}
\newcommand{\al}{\alpha}
\newcommand{\bt}{\beta}
\newcommand{\pr}{\partial}
\newcommand{\hs}{\hspace{5mm}}
\newcommand{\up}{\upsilon}
\newcommand{\dg}{\dagger}
\newcommand{\vphi}{\vec{\phi}}
\newcommand{\ve}{\varepsilon}
\newcommand{\acc}{\\[3mm]}
\newcommand{\dl}{\delta}
\newcommand{\grad}{\vec\partial}
\def\tablecap#1{\vskip 3mm \centerline{#1}\vskip 5mm}
\def\p#1{\partial_#1}
\def\DP#1#2{D_{#1}\phi^{#2}}
\def\dP#1#2{\partial_{#1}\phi^{#2}}
\def\xh{\hat x}
\def\mod#1{ \vert #1 \vert }
\def\chapter#1{\hbox{Introduction.}}
\def\Sin{\hbox{sin}}
\def\Cos{\hbox{cos}}
\def\Exp{\hbox{exp}}
\def\Ln{\hbox{ln}}
\def\Tan{\hbox{tan}}
\def\Cot{\hbox{cot}}
\def\Sinh{\hbox{sinh}}
\def\Cosh{\hbox{cosh}}
\def\Tanh{\hbox{tanh}}
\def\Asin{\hbox{asin}}
\def\Acos{\hbox{acos}}
\def\Atan{\hbox{atan}}
\def\Asinh{\hbox{asinh}}
\def\Acosh{\hbox{acosh}}
\def\Atanh{\hbox{atanh}}
\def\frac#1#2{{\textstyle{#1\over #2}}}

\newcommand{\Ref}[1]{(\ref{#1})}
\newcommand{\ie}{{\it i.e.}}
\newcommand{\cmod}[1]{ \vert #1 \vert ^2 }
\newcommand{\nhat}{\mbox{\boldmath$\hat n$}}
\nopagebreak[3]

\title{Solitons and deformed lattices I}
\author{
B. Hartmann\thanks{e-mail address: Betti.Hartmann@durham.ac.uk}
and
W.J. Zakrzewski\thanks{e-mail address: W.J.Zakrzewski@durham.ac.uk}
\\
\\
\\
Department of Mathematical Sciences,University of Durham, \\
Durham DH1 3LE, UK\\} \date{}

\maketitle
\begin{abstract}
We study a model describing some aspects of the dynamics of biopolymers. The
models involve
either one or two finite chains with a number $N$ of sites that represent the ``units"
of a biophysical system. The mechanical degrees of freedom of these
chains are coupled to the internal degrees of freedom through  position 
dependent excitation transfer functions. We reconsider the case of the one
chain model discussed by Mingaleev et al. and present new results concerning 
the soliton sector of this model.
We also give new (preliminary) results in 
the two chain model in which case we have introduced an 
interaction potential inspired by the Morse potential.

\end{abstract} 
\medskip

\section{Introduction}

Recently some work has been done on the study of curvature-induced
symmetry breaking effects in nonlinear Schr\"odinger models \cite{Gai}.
The idea here is to study a one-dimensional
discrete Nonlinear Schr\"odinger equation 
\begin{equation}
i{d\over dt}\psi_i\,+\,\sum_k J_{ik}\,\psi_k\,+\, \chi\vert \psi_i\vert\sp2
\psi_i\,=\,0
\end{equation}
in which the excitation transfer function ($J_{ik}$) depends
on the positions of the lattice 
points. Often \cite{Abl} one restricts oneself to the 
nearest neighbour approximation in which case 
$J_{ik}=J\delta_{i-k,\pm 1}$ but, as pointed out in
\cite{Gai}, a more general case involves $J_{ik}=J(\vert\vec{r_i}-\vec{r_k}
\vert)$ where $\vec{r_i}$ describes the spatial position of 
the $i\sp{th}$ lattice site. Of course we expect $J(\vert\vec{r_i}-\vec{r_k}
\vert)$ to be a fast decreasing function of its argument.
When all the points of the lattice  lie along a straight line, i.e. $\vec{r_i}
=\vec a + \alpha_i \vec b$, where $\alpha_i$ is a linearly growing
 function of $i$,
the results are not that different from the case of a regular lattice
with the nearest neighbour approximation.
However, as  pointed out in \cite{Gai}, any curvature in $\vec{r}_i$ can induce
extra effects which do affect the behaviour of the $\psi$ fields.

In \cite{Gai} the authors have studied the case when the lattice points lie on a
parabola with some curvature. Then they have considered $J(a)$ to be given
by $J(a)\,=\,J\,\exp{(-\alpha a)}$ and they have showed that the effective
Hamiltonian for the $\psi$ field involves a double well potential
whose shape depends on the curvature of the parabola.
This has an implication on the ground states of the theory.

This idea has then been applied to the study of the dynamics of biopolymers
\cite{Gai2}. The authors of \cite{Gai2} have considered the case 
of an excitation field $\psi$ 
on a one polymer chain. The links of the polymers have
been allowed to move, and their motion
has been controlled by their inter-link forces and 
also by the force due to the excitation field $\psi$.

The total Hamiltonian describing the system they have considered is given by
\begin{equation}
H=T+U+V
\end{equation}
where the kinetic energy $T$ and the inter-particle interaction
potential $U$ read:
\begin{equation}
T= \frac{M}{2}\sum_{i} (\frac{d\vec{r}_{i}}{dt})^2\quad \hbox{and}\quad
U(\vec{r}_{i})=U_{S}+U_{R}+U_{B}  \ .
\end{equation}
The streching energy $U_S$ is given by:
\begin{equation}
U_{S}=\frac{\sigma}{2}\sum_{i}
(|\vec{r}_{i}-\vec{r}_{i-1}|-a)^2 \ .
\end{equation}
$\sigma$ is the elastic module of the streching rigidity of the chain
and $a$ denotes the equilibrium lattice spacing. The repulsive potential 
\begin{equation}
U_{R}=\frac{\delta}{2}\sum_{i}\sum_{k\neq i}(d-
|\vec{r}_{i}-\vec{r}_{k}|)^2 \Theta(d-
|\vec{r}_{i}-\vec{r}_{k}|)
\end{equation}
comes only into play if the distance between two particles 
becomes smaller than the diameter $d$ of the particles themselves.
I.e. if the particles start to overlap, they are repelled.
However, since the strength $\delta$ of the repulsion is chosen finite,
particles are still allowed to overlap to some extend. 
The final bit of the inter-particle
interaction describes the bending energy:
\begin{equation}
U_{B}=\frac{\kappa}{2}\sum_{i}\frac{\theta_{i}^2}
{1-(\theta_{i}/\theta_{max})^2} \ , \ \theta_{i}=\theta_{i}(\vec{r}_{i}) \ .
\end{equation}
$\kappa$ is the elastic module of the bending rigidity while $\theta(\vec{r}_i)$ is
the angle between the two vectors $(\vec{r}_{i}-\vec{r}_{i-1})$
and $(\vec{r}_{i+1}-\vec{r}_{i})$ (for a detailed
formula see further in this paper or \cite{Gai2}). $\theta_{max}$ is the corresponding
maximal bending angle.
Finally  the potential of the complex scalar field $\psi$ reads:
\begin{equation}
V=\sum_{i}\{2|\psi_{i}|^2-\sum_{k\neq i} J_{ik}\psi_{i}^*
\psi_{k}-
\frac{1}{2}\chi |\psi_{i}|^4 \}
\end{equation}
where $\chi$ denotes the self-trapping nonlinearity and the exciation-transfer
coefficients are given by:
\begin{equation}
J_{ik}=(e^{\beta}-1)\exp(-\beta |\vec{r}_{i}-\vec{r}_{k}|).
\end{equation}
Note that the non-linear 
Schr\"odinger equation for the $\psi$ field 
and the Newton equations for the chain are coupled only by the $J_{ik}$ term.

In biophysical
systems the excitation $\psi_i$ can be thought of as an amide-I vibration, i.e. an excitation
in the C=0 bond of the peptide group, which transmits along the chain
and deforms it. Through this interaction
a new localized energy state is created, the Davydov soliton \cite{davy}, which can
transport energy without dispersion. Similarly, the electron excitation 
on a lattice leads to a localized state called the polaron through the electron-phonon
interaction \cite{davy,peyrard}.

One end of the polymer has been set to be free; the other one, assumed to be far away,
has been kept fixed.

The authors of \cite{Gai2} then have performed many interesting simulations 
of the equations which govern the dynamics of the electron and the lattice.
Their most spectacular results have involved them showing that a single excitation
of the $\psi$ field at the free end results, after a long period of time,
in a gradual folding of the chain. Then they have  explained their results as 
coming from the instability generated by the development  of the curvature
of the chain.

These exciting results have made us think of systems involving more chains
with an interchain interaction. 
Proteins often consist of two or more polypeptide chains and take on
characteristic structures in $3$-d space. Examples of such
secondary structures are the $\alpha$-helix and the $\beta$-sheet, where
in the latter pairs of chains lie side by side being stabilised by
hydrogen bonds between the carbonyl oxygen atom of one chain and 
the NH group of the other. Furthermore, DNA consists of two polynucleotide
chains which are connected to each other by hydrogen-bonding and are
coiled around each other in a double helix.

In view of these points, we have decided to extend the investigations
of \cite{Gai2} to systems of two chains with the simplest interchain interaction
as given in, e.g., \cite{Pey} and inspired by the Morse potential \cite{morse}.
So we have looked at a system of two chains with coupling
constants given by the appropriate
generalisations of those of \cite{Gai2}. Note that the coupling
 between the $\psi$ fields on the two chains allows the field to spread 
between them altering the values of the effective coupling constants
on each chain.

Our model is presented in the next section.

To test our program, we have, first of all,  tried to reproduce
the results of \cite{Gai2}. Unfortunately, ref \cite{Gai2} does not 
give all the details of their simulations - so in fact, 
not being sure what their initial conditions had been, we have not
been able to reproduce their results. However, we have found some interesting
properties of the system which we discuss in the following section.

Then we have looked at the system involving two chains. 
Our results are presented in Section 4.

We end the paper with some conclusions and our plans for the further studies.

\section{Our model}

As our model we take a straightforward generalisation of the model of 
\cite{Gai2} to which we have added an interaction between the chains
$W_{int}$.

\subsection{The Hamiltonian}
Thus the total Hamiltonian is given by~:
\begin{equation}
H=T+U+V+W_{int}
\end{equation}
with the kinetic energy
\begin{equation}
T=\sum_{j=1}^{2} \frac{M_{j}}{2}\sum_{i} (\frac{d\vec{r}_{ij}}{dt})^2 \ , \
\end{equation}
and the inter-particle interaction:
\begin{equation}
U(\vec{r}_{ij})=U_{S}+U_{B}+U_{R}
\end{equation}
where the stretching energy is a sum of the respective stretching energies
on chain $j=1$ and chain $j=2$:
\begin{equation}
U_{S}=\sum_{j=1}^{2}\frac{\sigma_{j}}{2}\sum_{i}
(|\vec{r}_{ij}-\vec{r}_{i-1,j}|-a_j)^2 \ .
\end{equation}
Similarly, the bending energy reads:
\begin{equation}
U_{B}=\sum_{j=1}^{2}\frac{\kappa_{j}}{2}\sum_{i}\frac{\theta_{ij}^2}
{1-(\theta_{ij}/\theta_{max})^2} \ , \ \theta_{ij}=\theta_{ij}(\vec{r}_{ij}) \ .
\end{equation}
Finally, the repulsive potential is a sum of the respective repulsive
potentials on the two chains and a new term, which describes the repulsion
if the two chains come closer than $d_{j'j}$ to each other:
\begin{eqnarray}
\label{ur}
U_{R}&=&\sum_{j=1}^{2}(\frac{\delta_{j}}{2}\sum_{i}\sum_{k\neq i}(d_j-
|\vec{r}_{ij}-\vec{r}_{kj}|)^2\Theta(d_j-
|\vec{r}_{ij}-\vec{r}_{kj}|) \nonumber \\
&+& \sum_{j^{'}\neq j}\frac{\delta_{j^{'}j}}{2}
\sum_{i} (d_{j^{'}j}-|\vec{r}_{ij}-\vec{r}_{ij^{'}}|)^2)
\Theta(d_{j^{'}j}-|\vec{r}_{ij}-\vec{r}_{ij^{'}}|) \ .
\end{eqnarray}  

 The energy of the excitation $\psi_{ij}$ is given by~:
\begin{equation}
\label{v}
V=\sum_{j=1}^{2}\sum_{i}\{2|\psi_{ij}|^2-\sum_{k\neq i} J^j_{ik}\psi_{ij}^*
\psi_{kj}-\sum_{j^{'}\neq j}\sum_{l} K^{jj^{'}}_{il}\psi_{ij}^*\psi_{lj^{'}}-
\frac{1}{2}\chi_j |\psi_{ij}|^4 \}
\end{equation}
with the energy transfer coefficients:
\begin{equation}
J^j_{ik}=\lambda_j(e^{\beta_j}-1)\exp(-\beta_j |\vec{r}_{ij}-\vec{r}_{kj}|) \ ,
\end{equation}
\begin{equation}
K^{jj^{'}}_{il}=(e^{\gamma}-1)\exp(-\gamma |\vec{r}_{ij}-\vec{r}_{lj^{'}}|) \ .
\end{equation}
The second term on the rhs of (\ref{v}) describes the energy transfer along one
chain, while the third term corresponds to the energy transfer between 
the two chains.

For the interaction potential between the chains we take:~
\begin{equation}
\label{morse}
W_{int}=\sum_{ j^{'}\neq j}\sum_{i} D \{ \exp(-\alpha |\vec{r}_{ij}-\vec{r}
_{ij^{'}}|)
-1 \}^2.
\end{equation}
The expression for the interaction potential is inspired by the Morse potential. The latter arises in  a convenient model for the potential in a diatomic molecule.
$D$ then is the potential energy for the bond formation and
$\alpha$ a parameter controlling the width of the potential well.
Since the minimum of (\ref{morse}) is given for $\vec{r}_{ij}=\vec{r}
_{ij^{'}}$ for all $i$  this potential leads to an attraction
 between the chains. 

Note that our terms are simple generalisations of the expressions from 
\cite{Gai2}. All fields and coupling constants have an extra index $j=1,2$
which tells us which chain they refer to.
We have also added a term $\sum_{j^{'}\neq j}\sum_{l} K^{jj^{'}}_{il}\psi_{ij}^*\psi_{lj^{'}}$ 
coupling $\psi_{ij}$ fields on two different chains and, as we have mentioned before,
$W_{int}$. In addition, we have multiplied $J^j_{ik}$ by an extra constant $\lambda_j$ to 
extend the model to cases where the extension of the interaction over the chain and the strength
of the interaction can be chosen independently from each other.

\subsection{Equations}
It is easy to derive equations which follow from our Hamiltonian. They are given
by:
\subsubsection{The Schr\"odinger equations}
\begin{eqnarray}
i \frac{\partial \psi_{ij}}{\partial t} &=& 
2\psi_{ij}-\sum_{k\neq i} (e^{\beta_j}-1)
\exp(-\beta_j |\vec{r}_{ij}-\vec{r}_{kj}|)\psi_{kj}\nonumber \\
&-& \sum_{j^{'}\neq j}\sum_{l} (e^{\gamma}-1)
\exp(-\gamma |\vec{r}_{ij}-\vec{r}_{lj^{'}}|)\psi_{lj^{'}}\nonumber\\
&-& \chi_j |\psi_{ij}|^2\psi_{ij}\ , \hspace{1cm}  j=1,2
\end{eqnarray}

and, for the chains themselves:

\subsubsection{The Newton equations}
\begin{eqnarray}
M_j\frac{d^2\vec{r}_{ij}}{dt^2} &=& 
-\nu_j\frac{d\vec{r}_{ij}}{dt} - \frac{dU}{d\vec{r}_{ij}}+ 
\sum_{m}\sum_{k\neq m}\frac{dJ^j_{mk}}{d\vec{r}_{ij}}\psi^{*}_{mj}\psi_{kj}
\nonumber \\
&+& \sum_{m}\sum_{j^{'}\neq j}\sum_{l}\frac{dK^{jj^{'}}_{ml}}{d\vec{r}_{ij}}\psi^{*}_{mj}\psi_{lj^{'}}
-\frac{dW_{int}}{d\vec{r}_{ij}}\ , \hspace{1cm}  j=1,2
\end{eqnarray} 
where $M_j$ is the mass of the particles on the $j$-th chain and $\nu_j$ are the
corresponding damping parameters.
The derivatives of the different potential terms with respect to $\vec{r}_{ij}$
read:
\begin{equation}
\frac{dU_{S}}{d\vec{r}_{ij}}=\sigma_j \left( (|\vec{r}_{ij}-\vec{r}_{i-1,j}|-a_j)
\cdot \frac{\vec{r}_{ij}-\vec{r}_{i-1,j}}{|\vec{r}_{ij}-\vec{r}_{i-1,j}|}-
(|\vec{r}_{i+1,j}-\vec{r}_{i,j}|-a_j)
\cdot \frac{\vec{r}_{i+1,j}-\vec{r}_{i,j}}{|\vec{r}_{i+1,j}-\vec{r}_{i,j}|} 
\right)
\end{equation}
\begin{equation}
\label{angle1}
\frac{dU_{B}}{d\vec{r}_{ij}}=\kappa_j\frac{\theta_{ij}\frac{d\theta_{ij}}
{d\vec{r}_{ij}}}{(1-(\theta_{ij}/\theta_{max})^2)^2}
\end{equation}
\begin{equation}
\frac{dU_{R}}{d\vec{r}_{ij}}=-\delta_j\sum_{k\neq i}
(d_j-|\vec{r}_{ij}-\vec{r}_{kj}|)
\cdot \frac{\vec{r}_{ij}-\vec{r}_{kj}}{|\vec{r}_{ij}-\vec{r}_{kj}|}
-\sum_{j^{'}\neq j}\delta_{j^{'}j}(d_{j^{'}j}-|\vec{r}_{ij}-\vec{r}_{ij^{'}}|)
\cdot \frac{\vec{r}_{ij}-\vec{r}_{ij^{'}}}{|\vec{r}_{ij}-\vec{r}_{ij^{'}}|}
\end{equation}
\begin{equation}
\sum_{m}\sum_{k\neq m}\frac{dJ^j_{mk}}{d\vec{r}_{ij}}
\psi^{*}_{mj}\psi_{kj}
= \lambda_j\beta_j (1-e^{\beta_j})\sum_{k\neq i} \exp 
\{-\beta_{j}|\vec{r}_{ij}-\vec{r}_{kj}|\}
\cdot \frac{\vec{r}_{ij}-\vec{r}_{kj}}{|\vec{r}_{ij}-\vec{r}_{kj}|}
\cdot (\psi^*_{ij}\psi_{kj}+\psi^*_{kj}\psi_{ij})  
\end{equation}
\begin{equation}
\sum_{m}\sum_{j^{'}\neq j}\sum_{l}\frac{dK^{jj^{'}}_{ml}}{d\vec{r}_{ij}}
\psi^{*}_{mj}\psi_{lj^{'}}
=\gamma(1-e^{\gamma})\sum_{j^{'}\neq j}\sum_{l}\exp(-\gamma|\vec{r}_{ij}-\vec{r}_{lj^{'}}|)
\cdot  \frac{\vec{r}_{ij}-\vec{r}_{lj^{'}}}{|\vec{r}_{ij}-\vec{r}_{lj^{'}}|}
\cdot (\psi^*_{ij}\psi_{lj^{'}}+\psi^*_{lj^{'}}\psi_{ij})
\end{equation}
\begin{equation}
\frac{dW_{int}}{d\vec{r}_{ij}}=-2D\alpha\sum_{j^{'}\neq j}
\{ \exp(-\alpha |\vec{r}_{ij}-\vec{r}
_{ij^{'}}|)-1 \}
\cdot\frac{\vec{r}_{ij}-\vec{r}_{ij^{'}}}{|\vec{r}_{ij}-\vec{r}_{ij^{'}}|}
\cdot \exp(-\alpha |\vec{r}_{ij}-\vec{r}_{ij^{'}}|).
\end{equation}
As in \cite{Gai2}, we have introduced a viscous damping $\nu_j$ resulting from the assumption of having an aqueous environment. This is a physical
reason for the introduction of such a term. On the other hand we will
try to find the lowest energy (stationary) state of our system and we will
use the absorption to reduce the energy and to `drive' the system to its
ground state. Thus, in practice, we will start with a good approximation 
to the ground (lowest energy) state and then evolve it with nonzero
absorption  - thus letting it `flow' towards this true ground state. 

To proceed further we need to determine the dependence of
 angles $\theta_{ij}$ on
$\vec{r}_{ij}$. Geometrical reasoning gives:
\begin{equation}
\theta_{ij}=\pi-\tilde{\theta}
\end{equation}
where
\begin{equation}
\tilde{\theta}=\arccos(\frac{\vec{r}_{ij}^2-\vec{r}_{i+1,j}
\cdot\vec{r}_{ij}-\vec{r}_{ij}
\cdot\vec{r}_{i-1,j}+\vec{r}_{i+1,j}
\cdot\vec{r}_{i-1,j}}
{|\vec{r}_{ij}-\vec{r}_{i-1,j}||\vec{r}_{i+1,j}-\vec{r}_{i,j}|}).
\end{equation}
Then
\begin{equation}
\frac{d\theta_{ij}}{d\vec{r}_{ij}}=[\frac{d\theta_{ij}}{d\vec{r}_{ij}}]_{i}
+[\frac{d\theta_{ij}}{d\vec{r}_{ij}}]_{i-1}+[\frac{d\theta_{ij}}{d\vec{r}_{ij}}]_{i+1}
\label{dphi}
\end{equation}
The first term on the rhs is given by:
\begin{equation}
[\frac{d\theta_{ij}}{d\vec{r}_{ij}}]_{i}=(1-
(\frac{\vec{r}_{ij}^2-\vec{r}_{i+1,j}
\cdot\vec{r}_{ij}-\vec{r}_{ij}
\cdot\vec{r}_{i-1,j}+\vec{r}_{i+1,j}
\cdot\vec{r}_{i-1,j}}
{|\vec{r}_{ij}-\vec{r}_{i-1,j}||\vec{r}_{i+1,j}-\vec{r}_{i,j}|})^{2})^{-1/2}
\cdot\frac{F_1(\vec{r}_{ij})}{F_2(\vec{r}_{ij})}
\end{equation}
with
\begin{eqnarray}
F_1(\vec{r}_{ij})&=&
(2\vec{r}_{ij}-\vec{r}_{i+1,j}-\vec{r}_{i-1,j})
(|\vec{r}_{ij}-\vec{r}_{i-1,j}||\vec{r}_{i+1,j}-\vec{r}_{ij}|)\ \nonumber\\
&-&(\vec{r}_{ij}^2-\vec{r}_{i+1,j}
\cdot\vec{r}_{ij}-\vec{r}_{ij}
\cdot\vec{r}_{i-1,j}+\vec{r}_{i+1,j}
\cdot\vec{r}_{i-1,j})\ \nonumber \\
&\cdot&((\vec{r}_{ij}-\vec{r}_{i-1,j})\frac{|\vec{r}_{i+1,j}
-\vec{r}_{i,j}|}{|\vec{r}_{i,j}-\vec{r}_{i-1,j}|}-
(\vec{r}_{i+1,j}-\vec{r}_{ij})\frac{|\vec{r}_{i,j}-
\vec{r}_{i-1,j}|}{|\vec{r}_{i+1,j}-\vec{r}_{i,j}|})
\end{eqnarray}
and
\begin{equation}
F_2(\vec{r}_{ij})=(|\vec{r}_{ij}-\vec{r}_{i-1,j}|
|\vec{r}_{i+1,j}-\vec{r}_{ij}|)^2.
\end{equation}

The second and the third terms on the rhs of (\ref{dphi})
are given by similar expressions
with functions $F_3$, $F_4$, and  $F_5$, $F_6$, respectively.

\subsubsection{New terms for the curvature and the torsion}
We have also considered different terms to describe the effects
of the curvatures of the chains. Thus we have replaced
$U_B$ by $U_C$:
\begin{equation}
U_C=\sum_{j=1}^{2}\Lambda^{'}_{j}\sum_{i} k_{ij}^{2}
\end{equation}
where $k_{ij}$ is the curvature of the $j$th  chain at site $i$ 
and so $U_C$ is given by
\begin{equation}
U_C=\sum_{j=1}^{2}\Lambda_{j}\sum_{i} (1-\frac{(\vec{r}_{ij}-\vec{r}_{i-1,j})
(\vec{r}_{i+1,j}-\vec{r}_{ij})}{|\vec{r}_{ij}-\vec{r}_{i-1,j}|
|\vec{r}_{i+1,j}-\vec{r}_{ij}|})
\end{equation}
where $\Lambda_{j}=2\Lambda^{'}_{j}$.\\
Replacing $\frac{dU_B}{d\vec{r}_{ij}}$ by $\frac{dU_C}{d\vec{r}_{ij}}$, we get:
\begin{equation}
\frac{dU_C}{d\vec{r}_{ij}}=\Lambda_{j}(\frac
{F_1(\vec{r}_{ij})}{F_2(\vec{r}_{ij})}+\frac
{F_3(\vec{r}_{ij})}{F_4(\vec{r}_{ij})}+\frac
{F_5(\vec{r}_{ij})}{F_6(\vec{r}_{ij})})
\end{equation}
with $F_1$ etc. given by the previous equations.\

Now, comparing with the expressions we have had before, we see that
\begin{equation}
k^2_{ij}\propto (1+\cos \tilde{\theta})=(1+\cos (\pi-\theta_{ij}))=
(1-\cos \theta_{ij})
\label{cur}
\end{equation}
so that for $\theta_{ij}=0$ (straight line), $k^2_{ij}=0$.

We can also introduce torsion. To do this we note that 
 from Frenet's formulas we have 
\begin{equation}
\tau^2=\frac{d\vec{n}}{ds}-\kappa^2=\frac{d^3 \vec{r}}{ds^3}-\kappa^2.
\end{equation}

Thus it would make sense to  use an analog of (\ref{cur}) and so write
\begin{equation}
\tau^2\propto(1-\cos \phi)
\end{equation}
where $\phi$ is the angle between the two planes we compare. For $\phi=0$ 
it is clear that
$\tau^2=0$. In general $\phi$ is given by:
\begin{equation}
\cos \phi=\frac{[(\vec{r}_{ij}-\vec{r}_{i-1,j})\times
(\vec{r}_{i+1,j}-\vec{r}_{ij})]\cdot[(\vec{r}_{i+1,j}-\vec{r}_{ij})\times
(\vec{r}_{i+2,j}-\vec{r}_{i+1,j})]}{\sin\theta_{ij}\sin\theta_{i+1,j}
|\vec{r}_{ij}-\vec{r}_{i-1,j}|
|\vec{r}_{i+1,j}-\vec{r}_{ij}|^{2}
|\vec{r}_{i+2,j}-\vec{r}_{i+1,j}|}.
\label{angle}
\end{equation}
This can be rewritten as 
\begin{eqnarray}
\cos \phi &=& \frac{[(\vec{r}_{ij}-\vec{r}_{i-1,j})
\cdot (\vec{r}_{i+1,j}-\vec{r}_{ij})]
[(\vec{r}_{i+1,j}-\vec{r}_{ij})
\cdot (\vec{r}_{i+2,j}-\vec{r}_{i+1,j})]}
{\sin\theta_{ij}\sin\theta_{i+1,j}
|\vec{r}_{ij}-\vec{r}_{i-1,j}|
|\vec{r}_{i+1,j}-\vec{r}_{ij}|^2
|\vec{r}_{i+2,j}-\vec{r}_{i+1,j}|}\   \nonumber \\
&-& \frac{[(\vec{r}_{ij}-\vec{r}_{i-1,j})
\cdot (\vec{r}_{i+2,j}-\vec{r}_{i+1,j})]
}{\sin\theta_{ij}\sin\theta_{i+1,j}
|\vec{r}_{ij}-\vec{r}_{i-1,j}|
|\vec{r}_{i+2,j}-\vec{r}_{i+1,j}|}.
\label{angle2}
\end{eqnarray}
We know that $\sin \theta_{ij}=\sin\tilde{\theta}_{ij}$ and that $\sin(\arccos(x))=
\sqrt{1-x^2}$. Thus 
\begin{equation}
\sin \theta_{ij}=\sqrt{1-(\frac{\vec{r}_{ij}^2-\vec{r}_{i+1,j}
\cdot\vec{r}_{ij}-\vec{r}_{ij}
\cdot\vec{r}_{i-1,j}+\vec{r}_{i+1,j}
\cdot\vec{r}_{i-1,j}}
{|\vec{r}_{ij}-\vec{r}_{i-1,j}||\vec{r}_{i+1,j}-\vec{r}_{i,j}|})^2}.
\end{equation}
For a straight line, we have to be careful since in Eq.(\ref{angle})
$\sin\theta_{ij}=0$. 

We intend to compare the effects of all these terms on the dynamics
of the chains and of the excitation field. The preliminary results indicate
that the dependence on the detailed form of these terms is not very strong.
So in this paper we report the results for the case of $U_{B}$ leaving the
detailed study of the dependence on the form of $U_C$ to a future
publication. The same applies to the effects
associated with non-vanishing torsion.
At the same time our observation of the weak dependence on the details
of $U_C$ shows that our results are quite generic in their nature
with most observed effects determined by the other terms in our
Hamiltonian (2).


\subsection{Initial and boundary conditions}
Next we have to decide on the initial and boundary conditions.

First of all we have chosen the chains, initially,  to lie parallel
to each other (and we have placed them along the $x$ axis in the $xy$ plane, starting
at $x=0$ and running to $x=N$, where $N$ is the total number of links
of each chain).
So we have put  $\vec{r}_{i1}|_{t=0}=(x_{i1},0,0)$, and 
$\vec{r}_{i2}|_{t=0}=(x_{i2},1,0)$. It can then be shown that all angles vanish identically:
\begin{equation}
\tilde{\theta}|_{t=0}=\arccos(-1)=\pi \rightarrow \theta_{ij}|_{t=0}=0.
\end{equation} 

It is possible to fix the ends of the chains ($x=N$) by requiring that
\begin{equation}
\frac{dx_{N,j}}{dt}=0 \ , \ \frac{dy_{N,j}}{dt}=0\ , 
\ \frac{dz_{N,j}}{dt}=0.
\end{equation}
However, in most of our simulations, the end has been allowed to move freely.

We cannot define $\theta_{ij}$  at $x=0$. Thus, at this point, we have taken
the bending angle as
the angle between the $x$-direction and the vector $\vec{r}_{2j}-\vec{r}_{1j}$:
\begin{equation}
\theta_{1j}=\arcsin(\sqrt{(y_{2j}-y_{1j})^2+
(z_{2j}-z_{1j})^2}/|\vec{r}_{2j}-\vec{r}_{1j}|).
\end{equation}

\section{One chain}
In this case we have put $\sigma_2=\kappa_2=\delta_2=\delta_{12}=D=
\beta_2=\gamma=\chi_2=\nu_2=0$. Thus the second chain is completely
decoupled and our problem is reduced to that of ref [3].

We have redone some of the work reported in \cite{Gai2}. We have started
the simulations by putting, initially,  $\psi_{1,1}=1$ and $\psi_{i,1}=0$, 
when $i=2,..N$.
Using the parameters of \cite{Gai2}, ie  $\sigma_1=1000$, $\kappa_1=0.06$,
$\delta_1=100$, $\beta_1=2$, $\chi_1=3.2$, $\nu_1=0.3$, $\theta_{max}=\pi/3$, $M_1=0.5$,
$d_1=0.6$ and in addition $\lambda_1=1$,
we have found that the links of the chain have moved very little, 
and that the motion was only in the $x$ direction. This is easy to understand 
as nothing breaks the symmetry keeping the motion restricted to the 
$y=z=0$ line. Of course, this symmetry can be broken dynamically
through the numerical inaccuracies but this process is extremely slow.

Hence we have broken the symmetry explicitly by introducing a small initial 
displacement of the chain in the $y$ as well as in the $z$ direction.
Thus we put initially $y_{1,1}=0.1$, $z_{1,1}=0.01$ and set all 
other $y_{i,1}$, $z_{i,1}$ equal to zero. In all our simulations,
the energy has decreased due to the absorption and has settled quite
 quickly to its final value.

We have first studied the existence of a soliton-like structure 
as a function of the parameters 
$\beta_1\equiv \beta$ and $\lambda_1$. We have found that the range of $\lambda_1$ for which the soliton doesn't get 
destroyed is very limited for all $\beta_1$. For $\lambda_1\geq 5$, $\psi_{i,1}$ is completely spread over the 
chain after some finite time, while for $\lambda_1=2,3$, a soliton-like structure is still seen to move up 
and down the chain, however, a lot of ``background" 
noise is present. A sharp soliton seems to be present
only for $\lambda_1\sim 1$.  We have thus fixed $\lambda_1\equiv 1$ and studied the dependence of the height,
 width and velocity of the soliton on the parameter $\beta_1\equiv\beta$. 
For the position $i_{max}$ and 
the height $(\psi_{i,1}\psi_{i,1}^{*})_{max}=\vert\psi_{i,1}\vert^2_{max}$ 
of the soliton's maximum, we have used a quadratic approximation.
For the width $\Delta$ of the soliton in the $x$-direction we can use two different definitions:
\begin{itemize}
\item a ``quantum-mechanical" definition: 
\begin{equation}
\Delta^{qm}= \sqrt{<x^2>-<x>^2} 
\end{equation}
with $ <x^2>=\frac{1}{N}\sum_{i}x_{i,1}^2\vert\psi_{i,1}\vert^2$ and
$<x>^2=(\frac{1}{N}\sum_i x_{i,1} \vert\psi_{i,1}\vert^2)^2$, or
\item a definition using the half-maximum of $\psi\psi^{*}$:  
\begin{equation}
\Delta^{1/2}=x_{1/2}^{+}-x_{1/2}^{-}
\end{equation}
where  $x_{1/2}^{+} > x_{max}$ and $x_{1/2}^{-} < x_{max}$ denote
the two $x$-values for which
$\vert\psi_{i,1}\vert\sp2(x_{1/2}^{+})=\vert\psi_{i,1}\vert\sp2(x_{1/2}^{-})
=(\vert\psi_{i,1}\vert\sp2)_{max}/2$.
\end{itemize} 

Our results for the height, position and width  
of the soliton as a function of $\beta_1\equiv\beta$ after $t=26.4$ sec are 
shown in Fig.s 1(a), (b), (c), (d) for three different values of the 
coupling $\chi_1\equiv\chi$. From these figures we see that
the soliton moves quicker with the decreasing of $\beta$.
Moreover, the height of the soliton's maximum decreases with
the decreasing of $\beta$, while at the same time the soliton gets
broader as can be seen from Fig.s 1(c) and 1(d).
Comparing Fig.s 1(c) and 1(d) it is also clear that the qualitative 
results from the 
two different expressions for the width are in good agreement with
each other, even though the ``quantum mechanical"
expression takes  into account all $\psi_{i,1}$ for
 $i \leq 40$ along the chain. 

For $\beta > 10$,
the excitation is completely trapped at its initial position
$\psi_{1,1}=1$. This can be explained by noting
that for large $\beta$, the $\psi$ field equation essentially
decouples from the chain positions. For $\beta < 1$ it spreads over the chain with an oscillating
behaviour of $\vert\psi_{i,1}\vert\sp2$. This is already noticeable for $\beta$s slightly 
larger 
than this value, where the soliton maximum decreases as it moves along the chain.
For the reasons above, we can only calculate the
maxima of the soliton etc. for the interval $\beta$ $\epsilon$ $[1:10]$.

It had already been noticed in \cite{Gai2} that the soliton exists only for $2.5 < \chi < 3.5$. We find the
same in our simulations with the initial excitation
being trapped at its initial position
for $\chi > 3.5$ and being completely destroyed for $\chi < 2.5$. Thus we have  chosen
$\chi=2.8$, $=3.0$ and $=3.2$ to investigate 
the dependence on this parameter. Clearly, as seen from
Fig. 1(a), the soliton moves faster as $\chi$ decreases. Moreover, its height 
decreases with decreasing $\chi$ (at least comparing the results for $\chi=3.2$ and $\chi=2.8$).
The curve for $\chi=3.0$ is not very conclusive; however, we believe that this is due to
the quadratic approximation we have used.
Both methods of computing the width of the soliton again give the same qualitative result
that the width decreases with increasing $\chi$. Thus the soliton is ``sharper" for larger values of 
$\chi$ as could have been expected.

Following \cite{lomdahl}, we have further studied the correlation between
the position of the soliton and the average displacement of the site $i$ given by~:
\begin{equation}
\delta_{i+1}-\delta_{i-1}=\sqrt{dx_{i+1,1}^2+dy_{i+1,1}^2+dz_{i+1,1}^2}-
\sqrt{dx_{i-1,1}^2+dy_{i-1,1}^2+dz_{i-1,1}^2}
\end{equation}
where $dx_{i,1}$, $dy_{i,1}$ and $dz_{i,1}$ are the displacements 
from the initial position
of the site $i$ in the $x$, $y$ and $z$ directions. For two
typical examples with $\chi=3.2$, $\beta=5$ and $\beta=10$, respectively,
with all other values given as before, we show $\delta_i$ as well as 
$\vert\psi_{i,1}\vert\sp2$ in Fig.s 2 (a) and 2 (b).
Clearly there is a correlation between these two quantities in the sense
that the maximum of $\vert\psi_{i,1}\vert\sp2$ is located in the region
where $\delta_i$ has a local minimum. 
Of course, the soliton movement generates also some ``background" noise
which influences the displacement of the sites from their initial position.
The minimum of the displacement at the left end of the chain
can be attributed to the fact that we have introduced an initial displacement of
the chain in both the $y$ and $z$ directions.

\section{Two chains}
We have performed several studies of a two chain system.
In each case we have originally placed two chains along the $x\ge 0$ axis, one of them at $y=0$, the other at $y=1$ (with $z=0$). Such a
 configuration does not depend on $z$ and this symmetry is preserved by the
equations of motion. Hence, to see any effects like folding,
 we have chosen to break the symmetry explicitly by
 displacing, initially, the first two links of the first chain in the 
$z$ direction by: $z_{1,1}=0.2$ and $z_{2,1}=0.1$. We have also used other values for this displacement but we have found our results not
 to be too sensitive to the initial values of this displacement (as long
as it was nonzero).

Then we have considered several initial conditions for the $\psi_{ij}$ fields.
In particular we have performed many simulations with $\psi_{12,1}=1$ (and all
others are zero), or with $\psi_{12,1}=0.95$, $\psi_{13,1}=i0.3$ and all other
vanishing.

We have performed these simulations for many values of $\beta_j$ and 
$\chi_j$. For the other parameters we have used $\sigma_1=\sigma_2=3000$,
$a_1=a_2=\kappa_1=\kappa_2=1$, $\delta_1=\delta_2=\delta_{12}=100$, 
$d_1=d_2=d_{12}=0.6$ and $\gamma=2,\, d=0.5,\, \alpha=1.8$.

At first this evolution is very fast, then it slows down and after a while 
it approaches the final configuration exponentially slowly. During the 
evolution at first the changes of the positions of the chains are clearly
visible, then become hardly noticeable and finally the chains look as
if they were essentially static.
The $\psi$ fields still evolve but even their evolution is not very 
dramatic.
Having reached this stage we are reasonably confident that our 
configurations are not very different from the final
asymptotic lowest energy configuration.
In fact, in each case, to be absolutely certain that we are well ``beyond''
any transient effects, we have run each of our simulations over
periods of several months of CPU time on various SUN workstations or PCs. 
Thus we are ready to present our results. They are somewhat qualitative.

We have preformed several simulations (for a range of values of
$\chi_j$ (from 2 to 9) and for various values of $\beta_j$ (from 1.5 to 2.5).

All the simulations have shown a localised (soliton-like) behaviour
 of the $\psi_{ij}$
fields, i.e. the effects on $\psi_{ij}$ fields were reasonably well localised. This localisation 
concerns both chains and it 
depends crucially on the values of $\chi_j$. For larger values (say $\chi_j=6.4$) 
we end up with a well defined soliton-like peak (clearly visible 
when you add the two chains) oscillating (relatively little) in size. 
Like in the case of a single chain the soliton moves and its speed depends
on $\beta_j$. For larger values of $\chi_j$ the soliton is narrower and
so more sharply defined while for smaller values it is more spread out.
For $\chi_j=3.2$ one can still identify a clearly localised structure
but whether this structure should be called a soliton remains to be decided.

We have found that the dynamics of the chains themselves depends
strongly on the soliton structure. For the case of small $\chi_j$ and thus
spread out $\vert\psi_{ij}\vert\sp2$, the chains get very much deformed 
(at least over short periods of time) in the sense
that they have large local curvature. For sharp solitons
however, i.e. large $\chi_j$, the behaviour of the chains is very similar
to that observed in the one chain model. Namely, the chains change their
shape very little and remain more or less straight.

We show a typical example of the minimal energy configuration
in Fig.~3. This is for $\beta_1=\beta_2=2.0$ and $\chi_1=\chi_2=4.0$.
 Note that 
this example corresponds
to a spread out $\vert\psi_{ij}\vert\sp2$. 
Clearly, the chains have large local curvature for
$0 \le x \le 60$. Looking at one chain by itself, it seems that
loops have formed. Especially the red-coloured chain has formed a large
one. Moreover, at small $x$, the two chains have overlapped each other
in the $x$-$y$-plane and it seems like they are winding around each other.
This winding also happens in the $\alpha$-helix, the feature that appears
e.g. in DNA.

Thus our results suggest that solitons exist when the chains are 
little deformed; while the large deformations of the chains correspond
to more spread out $\vert\psi_{ij}\vert\sp2$ fields. Moreover, we have not 
observed any behaviour which suggests any ``folding up of the chains".
At this stage our results are a little qualitative. We hope to present
a more quantative discussion of our results and an analysis  of the 
dependence on $\beta$ and $\kappa$ 
in the next paper (sometime in the future) \cite{hz}.


\section{Conclusions and Outlook}
In this paper we report our first results on  
the dynamics of biopolymer chains. While the authors of \cite{Gai2} 
put emphasis on the dynamics of a chain itself (they studied one chain)
and found that for suitable choices of the coupling constants 
the chain folds up, we have been mainly intersted in the dynamics
of the soliton moving along the chain. We have found that the soliton
exists only for very specific choices of the coupling constants. 
We have studied numerically
how the speed, height and width of the soliton depend on 
the coupling constants. Morever, we have confirmed the
existence of a direct 
correlation between the average displacement of the sites and the
location of the soliton.

For the case of the two chain model, we have introduced
suitable generalisations of the energy transfer coefficients
and the rigidity potential. Moreover, we have added an
attractive interaction potential
between the chains inspired by the Morse potential. We have found again
that the existence of the soliton depends crucially on the
non-linearity parameter $\chi$. However, while for one chain
the soliton is either completely spread over the chain or sharpely localised,
it seems that in the case of two chains there exist a sort of intermediate
situation. In this, a number of small peaks in $\vert\psi\vert\sp2$ exist that 
travel up and down the chain. Remarkably, the dynamics of the chains
themselves becomes only interesting in this latter case. We find features
like the formation of loops and the winding of the two chains around each
other which also happens in biological systems like e.g. DNA.
We plan to report a more detailed quantative analysis of the two-chain
model in a future publication \cite{hz}.  

\section{Acknowledgement}

WJZ wants to thank Y. Kivshar and S.F. Mingaleev for drawing his
attention to  this topic and showing him papers \cite{Gai} and \cite{Gai2}.\\
We also want to thank B. Piette, R. Ward, L. Brizhik and A. Eremko for their 
interest.\\
BH has been supported by an EPSRC grant.


\newpage
\begin{fixy}{0}
\begin{figure}
\centering
\epsfysize=13cm
\mbox{\epsffile{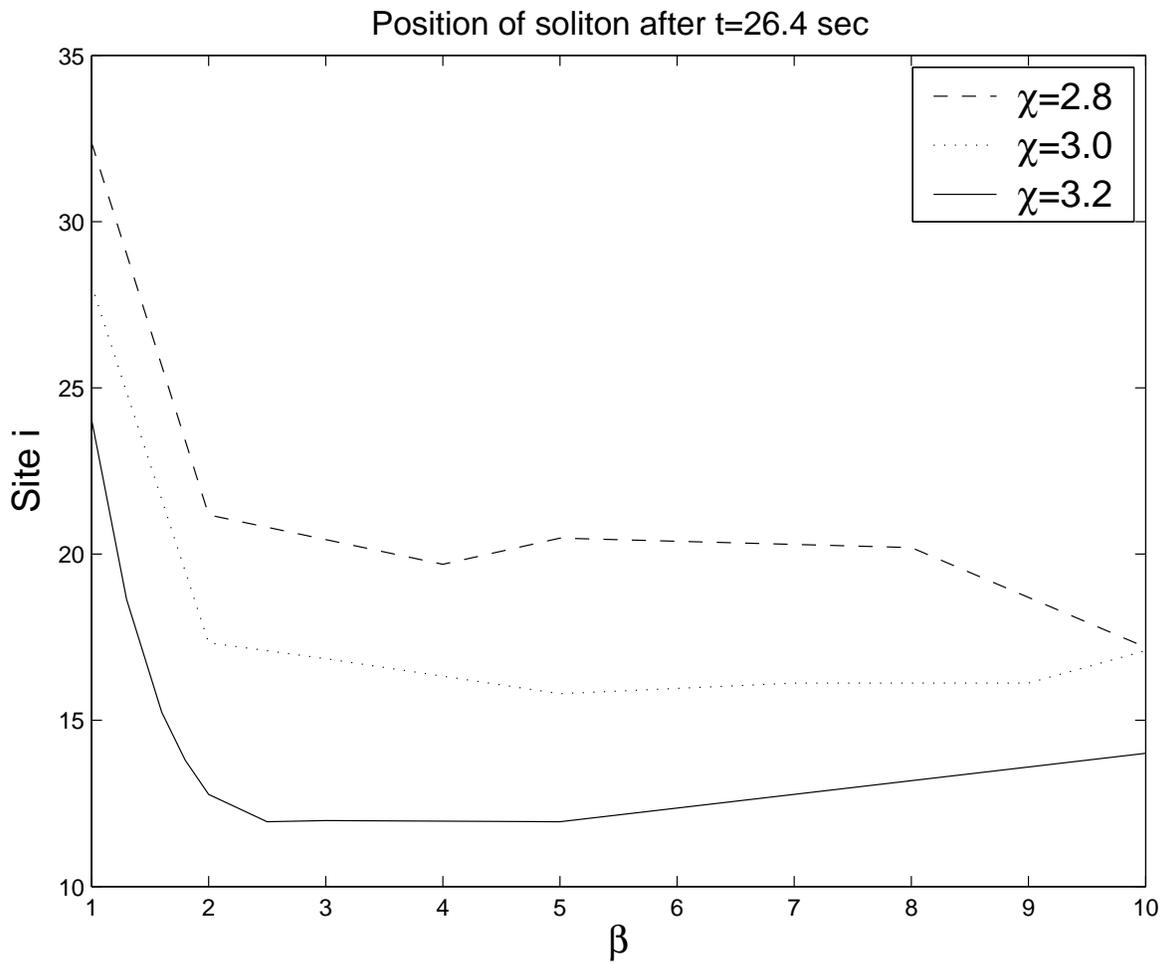}}
\caption{\label{Fig.1(a)} The dependence of the position of the soliton's
maximum at site $i$ is shown as function of $\beta_1\equiv\beta$  after 
$t=26.4$ sec for three different values of $\chi$.   }
\end{figure}

\begin{figure}
\centering
\epsfysize=13cm
\mbox{\epsffile{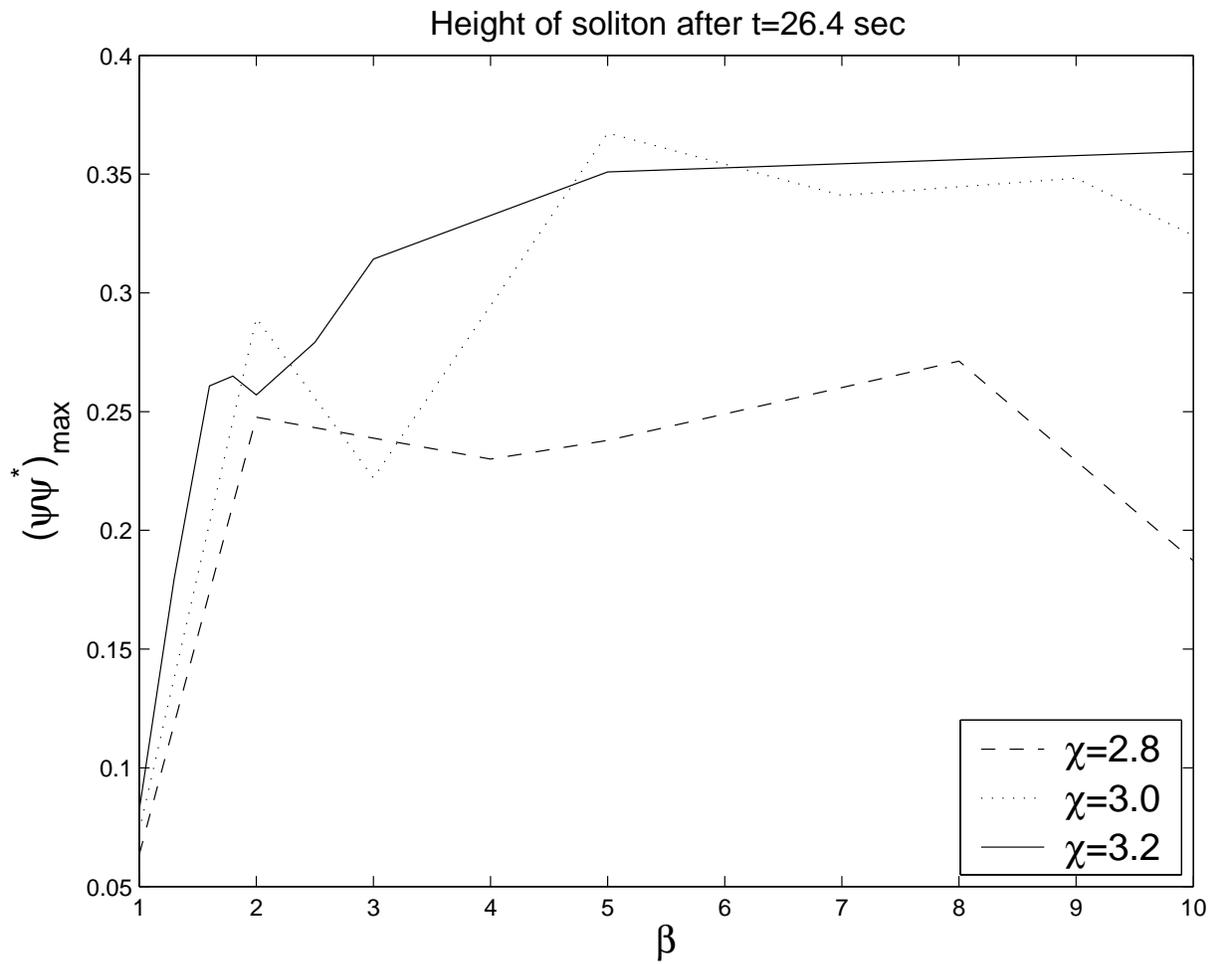}}
\caption{\label{Fig.1(b)}The height of the soliton's maximum
$(\psi\psi^{*})_{max}$ after $t=26.4$ sec is shown as function of 
$\beta_1\equiv\beta$ for three different values of $\chi$. }
\end{figure}

\begin{figure}
\centering
\epsfysize=13cm
\mbox{\epsffile{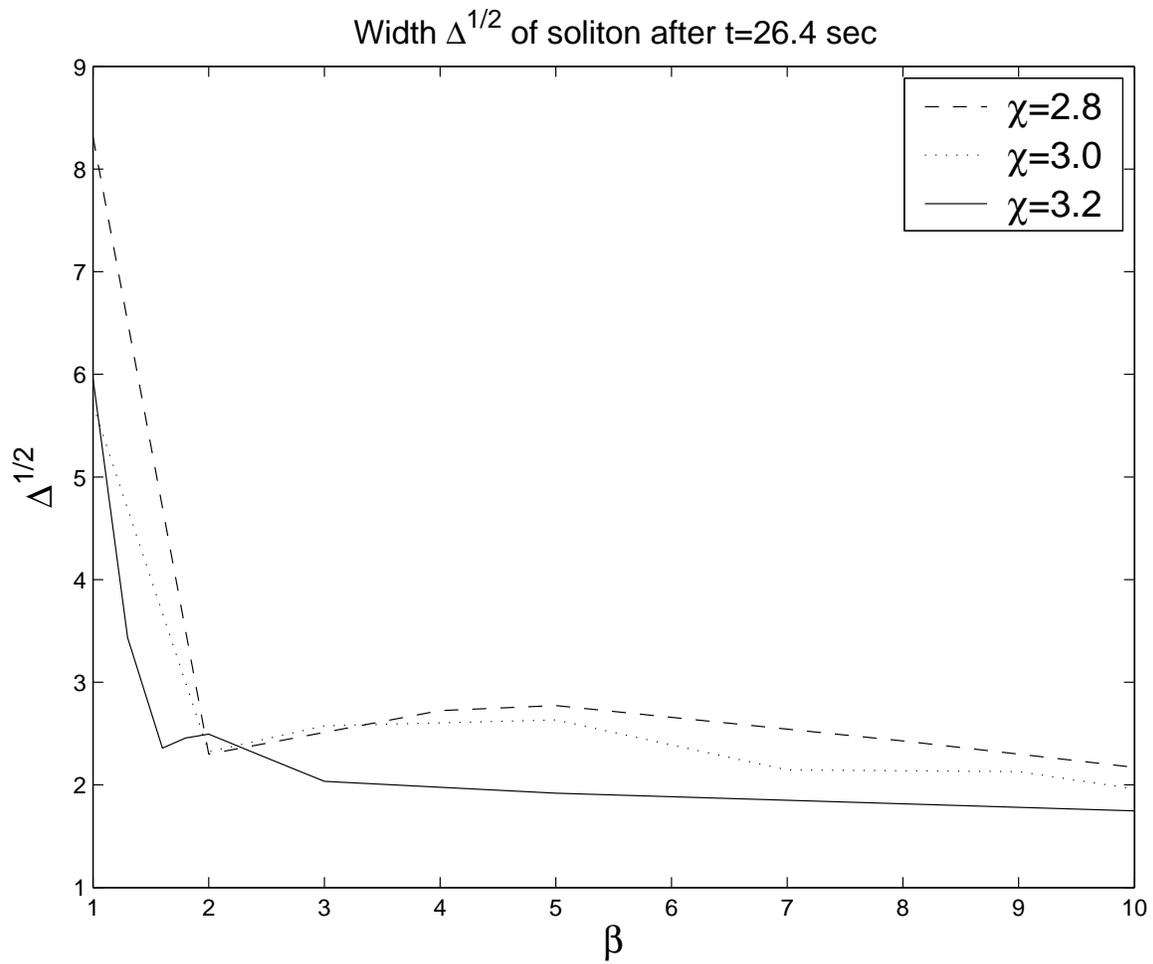}}
\caption{\label{Fig.1(c)}The ``half-maximum" width $\Delta ^{1/2}$ of 
the soliton after $t=26.4$ sec
is given as function of 
$\beta_1\equiv\beta$ for three different values of $\chi$.  }
\end{figure}
\begin{figure}
\centering
\epsfysize=13cm
\mbox{\epsffile{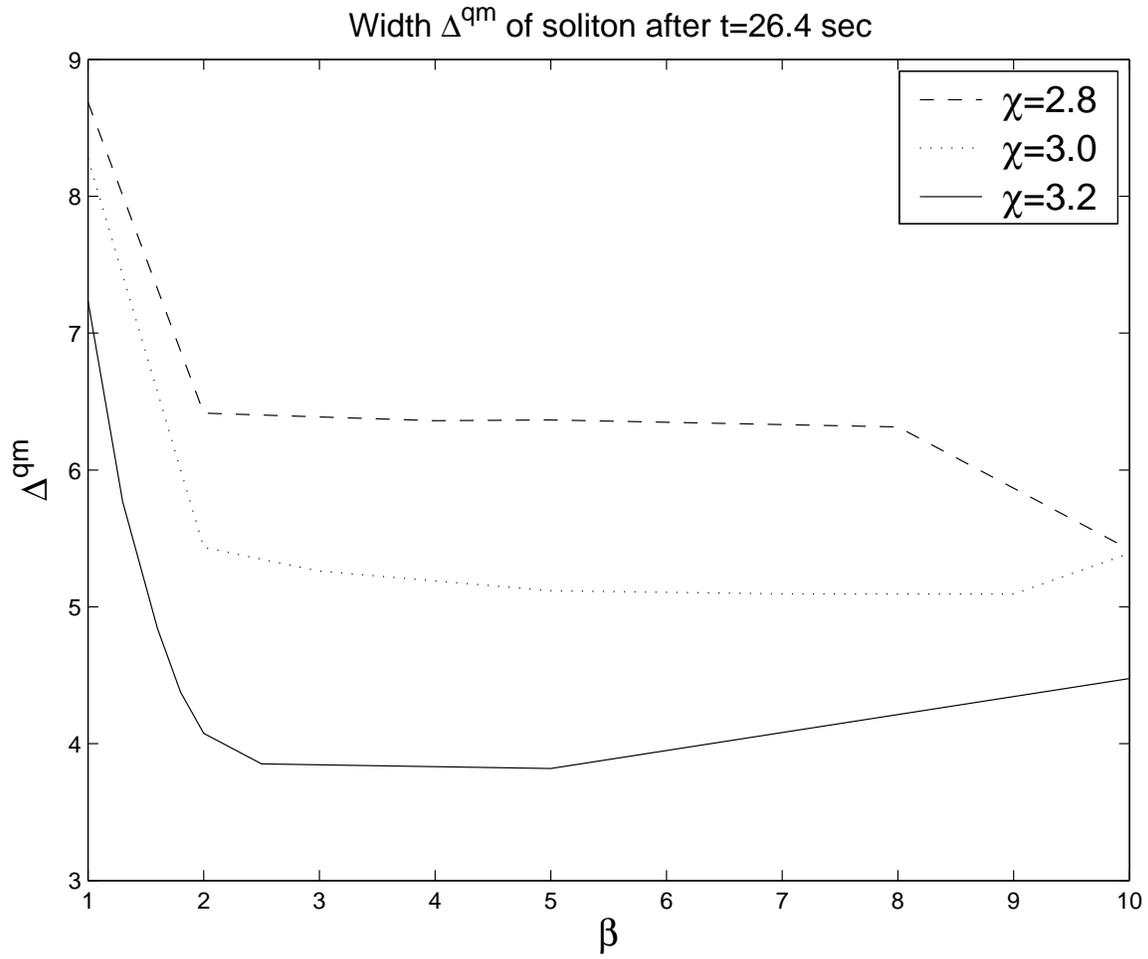}}
\caption{\label{Fig.1(d)}The ``quantum-mechanical" 
width $\Delta ^{qm}$ of the soliton after $t=26.4$ sec
is given as function of 
$\beta_1\equiv\beta$ for three different values of $\chi$.   }
\end{figure}
\end{fixy}
\begin{fixy}{0}
\begin{figure}
\centering
\epsfysize=13cm
\mbox{\epsffile{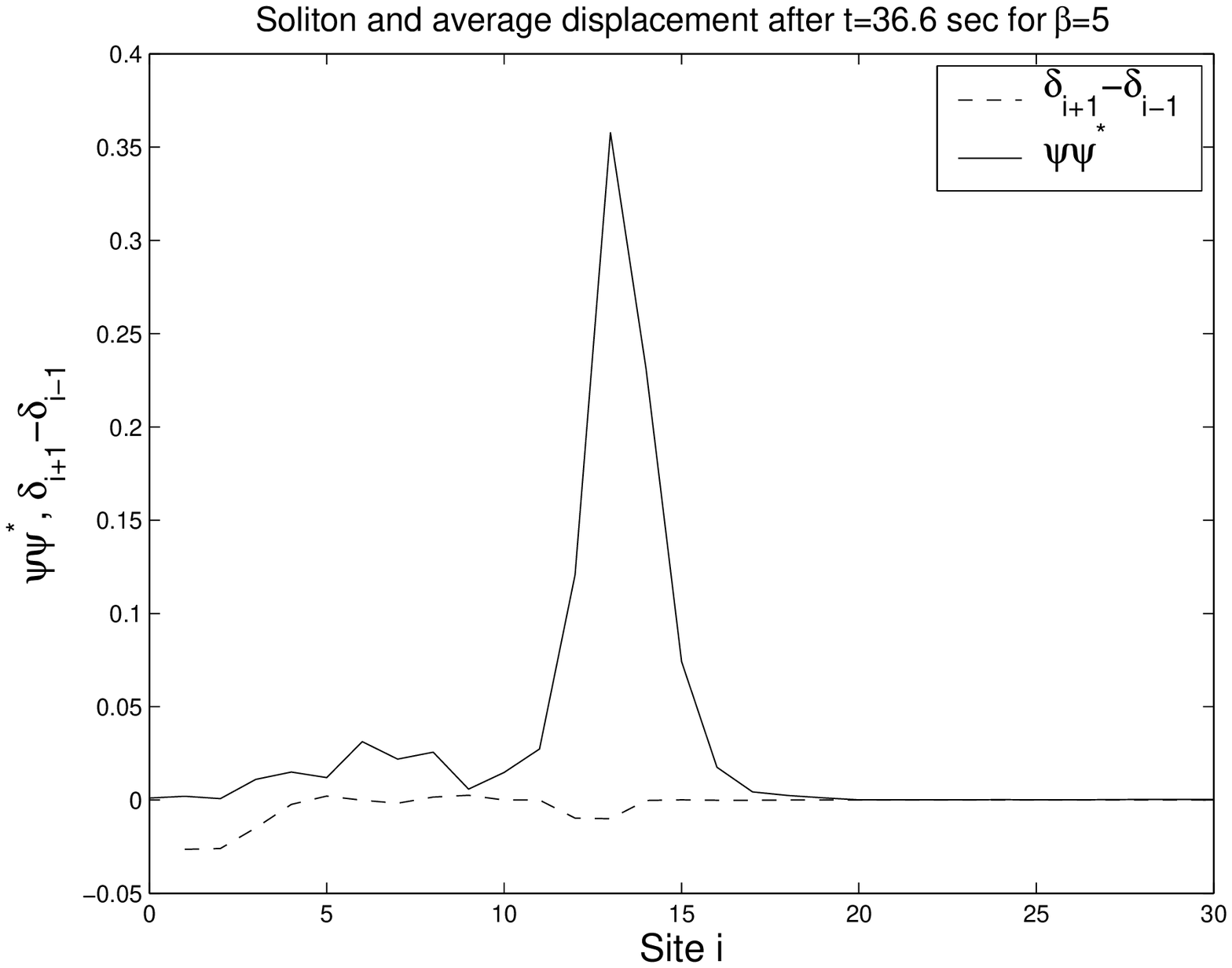}}
\caption{\label{Fig.2 (a)}The average displacement $\delta_i$ of the site 
$i$  as well
as $\psi_{i,1}\psi_{i,1}^*\equiv\psi\psi^*$ is shown as function of 
$i$ for $\beta=5$ after $t=36.6$ sec.  }
\end{figure}
\begin{figure}
\centering
\epsfysize=13cm
\mbox{\epsffile{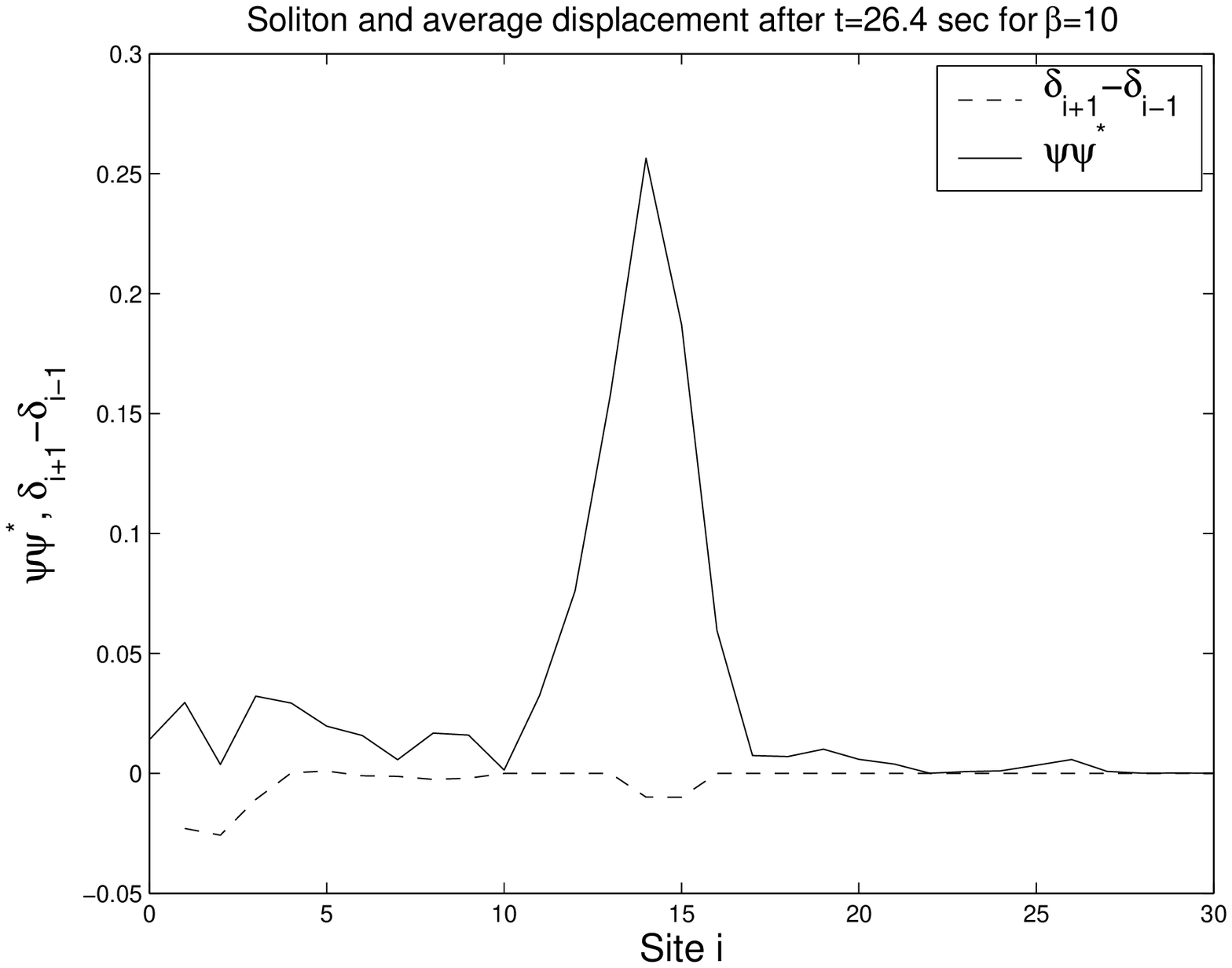}}
\caption{\label{Fig.2 (b)}The average displacement $\delta_i$ of the site $i$  as well
as $\psi_{i,1}\psi_{i,1}^*\equiv\psi\psi^*$ is shown as 
function of $i$ for $\beta=10$ after $t=24.6$ sec.  }
\end{figure}
\end{fixy}
\begin{fixy}{-1}
\begin{figure}
\centering
\epsfysize=13cm
\mbox{\epsffile{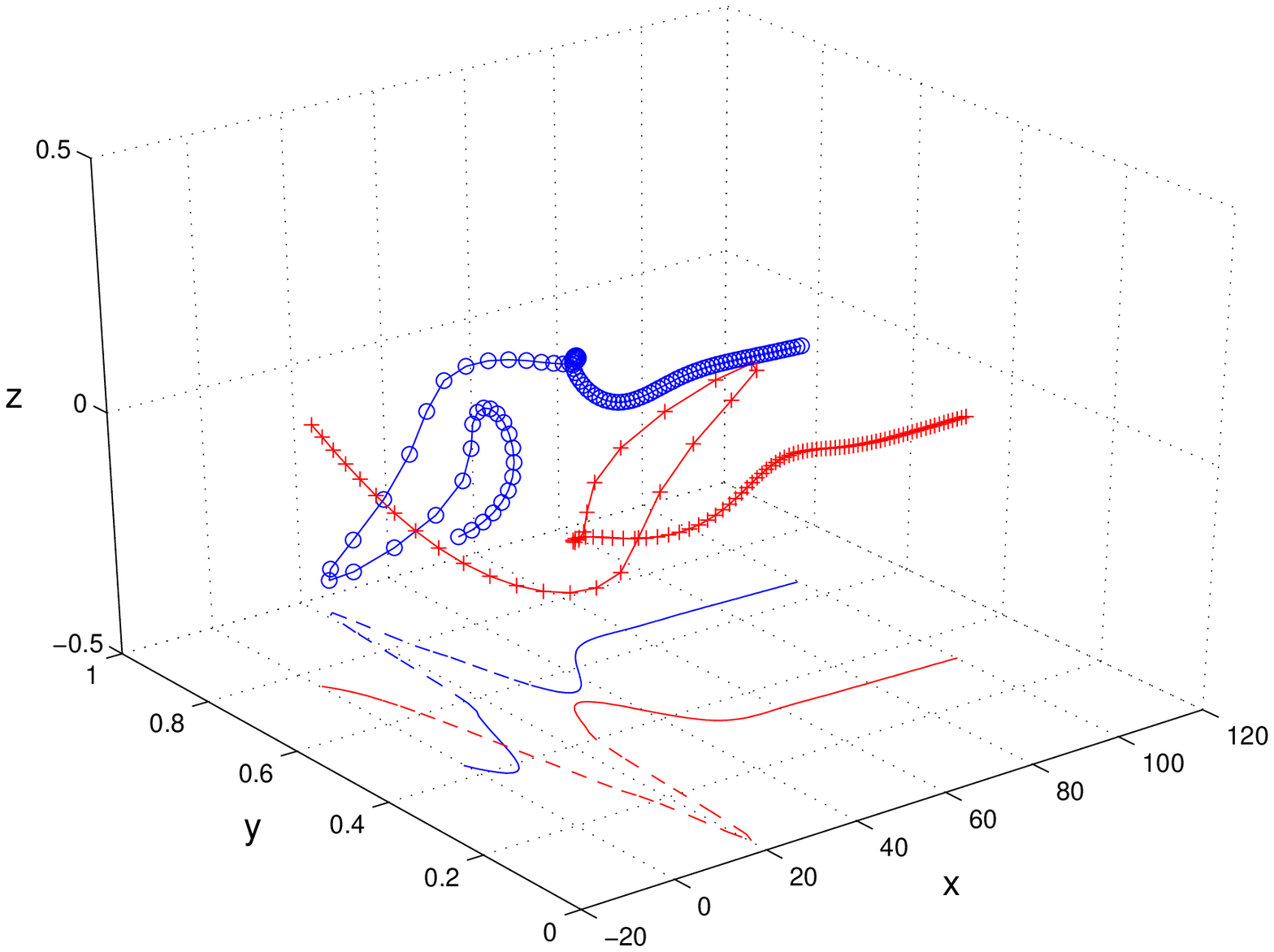}}
\caption{\label{Fig. 3} The minimal energy configuration of the two chains
for $\beta_1=\beta_2=2.0$ and $\chi_1=\chi_2=4.0$ is shown in $x$-$y$-$z$-space.
The two dashed lines at $z=-0.5$ represent the $x$-$y$-plots ($z\equiv 0$)
of the chains.   }
\end{figure}
\end{fixy}

\begin{thebibliography}{99}
\bibliographystyle{plain}
\bibitem{Gai} 
Yu. B. Gaididei, S. F. Mingaleev and P. J. Christiansen,
{\it Phys. Rev. E} {\bf 62}, 2000, pp. R53-56.
\bibitem{Abl}
{see eg M. J. Ablowitz and P. A. Clarkson, {\it Solitons, Nonlinear 
Evolution Equations and Inverse Scattering}, 1991, CUP.} 
\bibitem{Gai2}
{S. F. Mingaleev, Yu. B. Gaididei, P. J. Christiansen and
Yu. S. Kivshar, {\it Europhys. Lett.}{\bf 59}, 2002, pp. 403-409. }
\bibitem{davy} {A. S. Davydov, {\it Solitons in molecular systems}, Reidel, Dordrecht,
1985; A. Scott, {\it Phys. Rep. } {\bf 217}, 1992, pp. 1-67 }
\bibitem{peyrard} { {\it Nonlinear excitations in Biomolecules}, Ed.: M. Peyrard,
Springer, Berlin, 1996.}
\bibitem{Pey} 
{M. Peyrard and A. R. Bishop, {\it Phys. Rev. Lett.}{\bf 62}, 1989, pp. 2755-2758;
J. J. L. Ting and M. Peyrard, {\it Phys. Rev. E} {\bf 53}, 1996, pp. 1011-1020.}
\bibitem{morse}{P. M. Morse, {\it Phys. Rev.}  {\bf 34}, 1929, pp. 57-64.} 
\bibitem{lomdahl} P. S. Lomdahl, {\it Soliton models of protein dynamics}
in {\it Soliton theory: a survey of results}, Ed. A. P. Fordy, Manchester
University Press, 1990, pp. 209-232.
\bibitem{hz} B. Hartmann and W. J. Zakrzewski, 
{\it Solitons and deformed lattices II}, in preparation.



\end{thebibliography}
\end{document}